\documentclass{jpp}

\newcommand{\apj}{\mbox{\it Astrophysical Journal}}

\usepackage{color}
\usepackage{graphicx}
\usepackage{natbib}

\ifCUPmtlplainloaded \else
  \checkfont{msam10}
  \iffontfound
    \IfFileExists{amssymb.sty}
      {\typeout{^^JFound AMS Symbol fonts on the system, using the
                'amssymb' package.^^J}%
       \usepackage{amssymb}%
         
         \let\geq=\geqslant
      }{}
  \fi
\fi

\ifCUPmtlplainloaded \else
  \IfFileExists{amsbsy.sty}
    {\typeout{^^JFound the 'amsbsy' package on the system, using it.^^J}%
     \usepackage{amsbsy}}
    {}
\fi

\newsavebox{\astrutbox}
\sbox{\astrutbox}{\rule[-5pt]{0pt}{20pt}}

\title[Bridging the gap between collisional and collisionless shock waves]{Bridging the gap between collisional and collisionless shock waves}

\author[A. Bret and A. Pe'er]%
{Antoine  Bret$^{1,2}$, and Asaf Pe'er$^3$%
  \thanks{Email address for correspondence: antoineclaude.bret@uclm.es}
}

\affiliation{$^1$ETSI Industriales, Universidad de Castilla-La Mancha, 13071 Ciudad Real, Spain\\[\affilskip]
$^2$Instituto de Investigaciones Energ\'{e}ticas y Aplicaciones Industriales, Campus Universitario de Ciudad Real, 13071 Ciudad Real, Spain\\[\affilskip]
$^3$Department of Physics, Bar-Ilan University, Ramat-Gan, 52900, Israel
}

\date{?; revised ?; accepted ?. - To be entered by editorial office}

\begin{document}

\maketitle

\begin{abstract}
While the front of a fluid shock is a few mean-free-paths thick, the
front of a collisionless shock can be orders of magnitude thinner. By
bridging between a collisional and a collisionless formalism, we
assess the transition between these two regimes. We consider
non-relativistic, un-magnetized, planar shocks in electron/ion
plasmas. In addition, our treatment of the collisionless regime is
restricted to high Mach number electrostatic shocks. We find that the
transition can be parameterized by the upstream plasma parameter
$\Lambda$ which measures the coupling of the upstream medium. For
$\Lambda \lesssim 1.12$, the upstream is collisional, i.e. strongly coupled,
and the strong shock front is about $\mathcal{M}_1
\lambda_{\mathrm{mfp},1}$ thick, where $\lambda_{\mathrm{mfp},1}$ and
$\mathcal{M}_1 $ are the upstream mean-free-path and Mach number
respectively. A transition occurs for $\Lambda \sim 1.12$ beyond
which the front is $\sim
\mathcal{M}_1\lambda_{\mathrm{mfp},1}\ln \Lambda/\Lambda$ thick for
$\Lambda\gtrsim 1.12$. Considering $\Lambda$ can reach billions in
astrophysical settings, this allows to understand how the front of a
collisionless shock can be orders of magnitude smaller than the
mean-free-path, and how physics transitions continuously between these
2 extremes.
\end{abstract}

\maketitle

\section{Introduction}

Shock waves are very common in systems that involve fluid flows. Such systems occur on very different scales - from the microphysical scale to astronomical scales. As such, the properties of the shocks can vary considerably, depending on the environment. From the microphysical point of view, it is useful to discriminate between collisional and collisionless shocks. Collisional shock waves, first discovered in the $19^{th}$ century \citep{Salas2007}, can occur in any fluid as the result of a steepening of a large amplitude sound wave, or collision of two media \citep{Zeldovich}. The front of a collisional shock is necessarily at least a few mean-free-paths thick, as the dissipation from the downstream to the upstream occurs via binary collisions.

Collisionless shock waves were discovered later and can only form in plasma \citep{Petschek1958,Buneman1964,Sagdeev66}. The dissipation is provided by collective plasma phenomena instead of binary collisions. As a result, the front of such shocks can be orders of magnitude thinner than the mean-free-path. For example, the front of the bow shock of the earth magnetosphere in the solar wind is some 100 km thick \citep{PRLBow1,PRLBow2}. Yet, the proton mean-free-path at this location is about the Sun-Earth distance, nearly 7 orders of magnitude longer. Hence, if the earth bow shock were collisional, its front would be about 1 a.u. thick (see also \cite{balogh2013} \S2.1.3 and references therein).

Is it possible to bridge between these two regimes? How does a shock switches from a regime where its front is a few mean-free-paths thick, to another regime where its front is million times smaller? Exploring the intermediate case, bridging between collisional and collisionless shocks, is the aim of this article.

On the collisionless side, shock-accelerated particles which can enhance the density jump, or external magnetization which can reduce it, will be ignored \citep{Berezhko1999,BretJPP2018,BretPoP2019,BretApJ2020}.

The method implemented is explained in Section \ref{sec:method}. The big picture is as follows: we first present an evaluation of the shock front thickness in the collisional regime, then in the collisionless regime. The first task is achieved in Section \ref{sec:coll} using the Mott-Smith \emph{ansatz} \citep{MottSmith1951}, which writes the distribution function at any place along the shock profile, as a linear combination of the upstream and downstream Maxwellians.  We then follow \cite{Tidman1967} in Section \ref{sec:coll-less} for the collisionless case before we bridge between the 2 expressions of the front thickness in Section \ref{sec:bridge}, to propose an expression of the front thickness valid from the collisional to the collisionless regime.

\begin{figure}
\begin{center}
 \includegraphics[width=\textwidth]{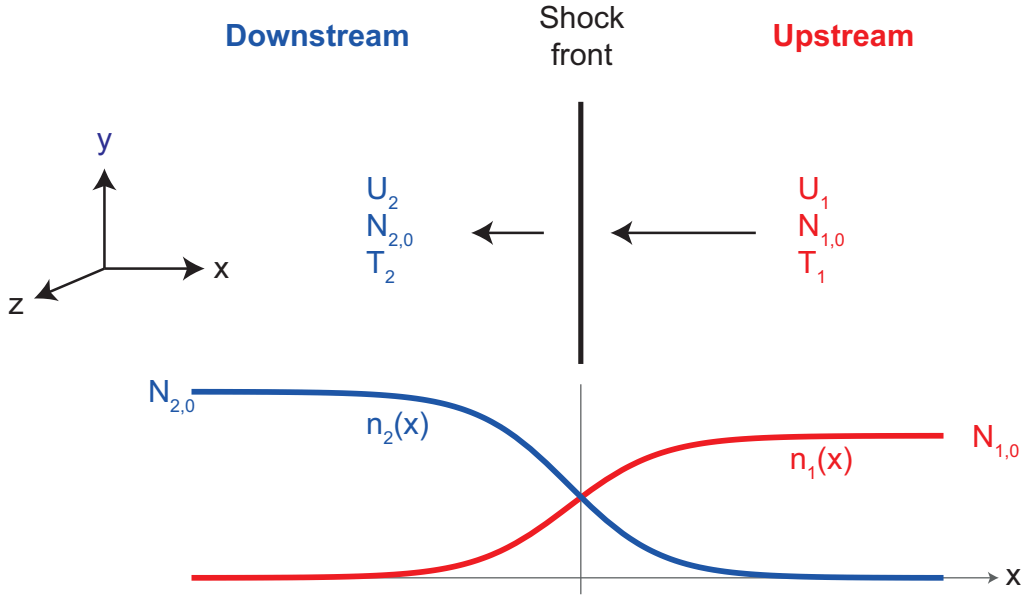}
\end{center}
\caption{Setup and notations.}\label{setup}
\end{figure}

\section{Method}\label{sec:method}
As previously stated, a fluid shock is mediated by collisions while a collisionless shock is mediated by collective effects.  For a plasma where only electrostatic fields are active (such is the case for an electrostatic shock, the kinetic equation accounting for both kinds of effects would formally read (\cite{Kulsrud2005}, p. 9),
\begin{equation}\label{eq:kujl}
\frac{\partial F}{\partial t} + \mathbf{v} \cdot \frac{\partial F}{\partial \mathbf{r}} + \frac{q \mathbf{E}}{m}\cdot\frac{\partial F}{\partial \mathbf{v}}
 =  \left( \frac{\partial F}{\partial t}\right)_c + \left( \frac{\partial F}{\partial t}\right)_w,
\end{equation}
where $q$ and $m$ are the charge and the mass of the species considered. The first term of the right-hand-side, namely $(\partial F/\partial t)_c$,  stands for the rate of change of the  distribution $F$ due to collisions. It is typically given by the Fokker-Planck operator. The second term, $(\partial F/\partial t)_w$, accounts for the effects of the waves and is given, for example, by the quasi-linear operator. In principle, accounting at once for these two collision terms with appropriate collision operators, should allow to describe a shock wave from the collisional to the fully collisionless regimes.

Resolving the shock front requires a formalism capable of resolving the entire shock profile. This is a notoriously difficult problem which has been greatly aided by the introduction of the so-called Mott-Smith \emph{ansatz} \citep{MottSmith1951}. Initially introduced for a neutral fluid, this \emph{ansatz} consisted in approximating the molecular distribution function $F$ along the shock profile by a linear combination of the upstream and downstream drifting Maxwellians,
\begin{eqnarray}\label{eq:amsatz}
F(\mathbf{v}) = &n_1(x)\left(\frac{m}{2\pi k_BT_1}\right)^{3/2}\exp\left(-\frac{m}{2 k_BT_1}(\mathbf{v}-\mathbf{U}_1)^2\right) \nonumber\\
                  &+ ~ n_2(x)\left(\frac{m}{2\pi k_BT_2}\right)^{3/2}\exp\left(-\frac{m}{2 k_BT_2}(\mathbf{v}-\mathbf{U}_2)^2\right),
\end{eqnarray}
 where $T_{1,2}$ and $\mathbf{U}_{1,2}$ are the upstream (subscript 1) and downstream (subscript 2)  temperatures and velocities respectively, determined by the Rankine-Hugoniot (RH) jump conditions (see figure \ref{setup})\footnote{All temperatures are not always considered constant in the main articles cited here \citep{MottSmith1951,Tidman1958,Tidman1967}. Yet, they are considered
so when it comes to computing the shock profile.}. The boundary conditions for the functions $n_{1,2}(x)$ are,
\begin{eqnarray}\label{eq:boudary}
 n_1(+\infty) &= N_{1,0} ~~~~  &n_1(-\infty) = 0, \nonumber\\
 n_2(+\infty) &= 0 ~~~~        &n_2(-\infty) = N_{2,0} ,
\end{eqnarray}
where again $N_{1,0}$ and $N_{2,0}$ fulfill the RH jump conditions.

Taking then the appropriate moments of the dispersion equation gives a differential equation which allows to determine the respective weights of the 2 Maxwellians in terms of $x$, hence the shock profile together with its front thickness \citep{MottSmith1951}.

The method implemented here consists in dealing with the collisional and the collisionless regimes separately.

\begin{itemize}
  \item We study the \emph{collisional regime} in Section \ref{sec:coll}. There we apply the Mott-Smith \emph{ansatz} using the BGK collision term \citep{Krook1954} as a collision operator for $(\partial F/\partial t)_c$ in Eq. (\ref{eq:kujl}), with $(\partial F/\partial t)_w=0$. Notably, \cite{Krook1954} presented 4 different collision operators through their Eqs. (3, 4, 5-6, 15-19). Those given by Eqs. (3, 4, 5-6), like $\nu(f-f_0)$\footnote{Here $\nu$ is a collision frequency, $f$ the distribution function and $f_0$ the equilibrium distribution function.}, have been widely used although they do not conserve all 3 quantities: particle number and/or momentum and/or energy. In \cite{Krook1954}, only the operator of Eqs. (15-19) does conserve all 3, hence  this is the one used here.

      \cite{Tidman1958} used the Fokker-Planck operator to deal with the problem, considering Eq. (\ref{eq:kujl}) with $(\partial F/\partial t)_w=0$  and,
\begin{equation}\label{GammaTid}
 \left( \frac{\partial F}{\partial t}\right)_c = \frac{4\pi e^4}{m_i^2}  \ln \Lambda \left( \frac{\partial F}{\partial t}\right)_{c,FP},
\end{equation}
where $(\partial F/\partial t)_{c,FP}$ is the Fokker-Planck collision operator, $m_i$ the ion mass, and $\Lambda$  the number of particles in the Debye sphere, that is, the co-called ``plasma parameter'' which measures the coupling of the plasma. As we shall see in Section \ref{sec:tid58}, the present treatment provides a more adequate bridging to the collisionless regime than \cite{Tidman1958}'s Fokker-Planck result.
  \item For the \emph{collisionless regime} we follow  in Section \ref{sec:coll-less} the collisionless result of \cite{Tidman1967} who also used the  Mott-Smith \emph{ansatz}. In recent years, the correctness of this approximation, namely that the distribution function is well approximated by superimposed drifting Maxwellians,  was validated numerically using Particle-In-Cell simulations \citep{Spitkovsky2008a}. \cite{Tidman1967} considered Eq. (\ref{eq:kujl}) with $(\partial F/\partial t)_c=0$, describing $(\partial F/\partial t)_w=0$ by the quasi-linear operator.
\end{itemize}

Having assessed the width of the shock front in the collisional and the collisionless regimes, we then bridge between the 2 expressions of the shock thickness in Section \ref{sec:bridge}.

\section{Collisional regime: applying \cite{Krook1954} collision term to the Mott-Smith \emph{ansatz}}\label{sec:coll}
We  switch to the reference frame of the shock and assume steady state in this frame. As in \cite{Tidman1958}, we consider  the distributions are functions of $(v_x,v_y,v_z)$ and assume quantities only vary with the $x$ coordinate. We therefore set  $\partial_{y,z,t}=0$ and $E_{y,z}=0$ so that equation (\ref{eq:kujl}), with $(\partial F/\partial t)_w=0$, reads for the ion distribution $F$,
\begin{equation}\label{eq:Krook3}
 v_x\frac{\partial F}{\partial x } + \frac{e E_x}{m_i}\frac{\partial F}{\partial v_x } = \left( \frac{\partial F}{\partial t }\right)_{c,BGK},
\end{equation}
where $m_i$ is the ion mass. The BGK collision term now reads \citep{Krook1954,Krook1956},
\begin{equation}\label{eq:KrookColl}
 \left( \frac{\partial F}{\partial t }\right)_{c,BGK} = \frac{1}{\sigma}\left( N^2 \Phi-N F \right),
\end{equation}
which vanishes for a Maxwellian distribution. According to \cite{Krook1954}, $``\sigma^{-1}$ $\times$ a density'' is a collision frequency $\nu$. In the present setting we define,
\begin{equation}\label{eq:nuBGK}
\frac{N_{1,0}}{\sigma} = \nu \sim \frac{v_{\mathrm{thi},1}}{\lambda_{\mathrm{mfp},1}} ~~ \Rightarrow  ~~  \sigma = \frac{N_{1,0} \lambda_{\mathrm{mfp},1}}{v_{\mathrm{thi},1}}  ,
\end{equation}
where $v_{\mathrm{thi},1}$ and $\lambda_{\mathrm{mfp},1}$ are the upstream thermal velocity and mean-free-path respectively. Then $N$ and $\Phi$ are given by Eqs. (15-19) of \cite{Krook1954},
\begin{eqnarray}
N                       &=& \int F d^3v =n_1 (x) + n_2(x),  \label{krookN}\\
\Phi               &=& \left( \frac{m_i}{2\pi k_BT } \right)^{3/2}\exp\left( -\frac{m_i}{2  k_BT }(\mathbf{v}-\mathbf{q} )^2 \right), \label{krookPhi} \\
\mathbf{q}         &=& \frac{1}{N}\int \mathbf{v} F d^3v, \label{krookq}\\
\frac{3  k_BT }{m_i} &=& \frac{1}{N}\int (\mathbf{v}-\mathbf{q} )^2 F d^3v,  \label{krookT}
\end{eqnarray}
 where $F$ has been considered of the form (\ref{eq:amsatz}). Multiplying equation (\ref{eq:Krook3}) by $v_y^2$ and integrating over $d^3v$ (see detailed calculation reported in Appendix \ref{ap:proofBGK}) gives an exact, simple result,
\begin{equation}\label{eq:CollOK}
 U_1\frac{k_BT_1}{m_i} ~ \frac{\partial n_1}{\partial x} +  U_2\frac{k_BT_2}{m_i}  ~ \frac{\partial n_2}{\partial x}
                 =  \frac{1}{3\sigma } (U_1-U_2 )^2  ~  n_1 n_2.
\end{equation}

This differential equation is structurally identical to the ones found in \cite{MottSmith1951,Tidman1958}. We show in Appendix \ref{ap:profile} how it yields density profiles like the ones pictured in Figure \ref{setup}, of the form,
\begin{eqnarray}\label{eq:profile}
n_1(x) &=& N_{1,0} \frac{1}{1+e^{-x/\ell}}, \nonumber \\
n_2(x) &=& N_{2,0} \frac{e^{-x/\ell}}{1+e^{-x/\ell}},
 \end{eqnarray}
implicitly defining the shock width $\ell$. From Eq. (\ref{eq:pattern})  we find the thickness of the shock according to the present formalism,
\begin{equation}\label{eq:ell_coll}
\ell=3\sigma \frac{  k_B(T_1 - T_2) U_2  }{ N_{1,0}  m_i (U_1-U_2)^2} = 3 \frac{N_{1,0} \lambda_{\mathrm{mfp},1}}{v_{\mathrm{thi},1}} \frac{  k_B(T_1 - T_2) U_2  }{ N_{1,0}  m_i (U_1-U_2)^2}.
\end{equation}

It is now convenient to use the RH jump conditions to express $\ell$ in terms of the upstream quantities, like the upstream Mach number and mean-free-path. The calculations reported in Appendix \ref{ap:RH} give,
\begin{eqnarray}\label{eq:stage}
\ell &=&  \lambda_{\mathrm{mfp},1}\frac{U_1}{v_{\mathrm{thi},1}}  \frac{ (\mathcal{M}_1^2+3 )  (5 \mathcal{M}_1^2  (\mathcal{M}_1^2+2 )-3 )}{5 \mathcal{M}_1^2  (\mathcal{M}_1^2-1)^2}, \nonumber \\
     &=&  \lambda_{\mathrm{mfp},1}   \frac{ (\mathcal{M}_1^2+3 )  (5 \mathcal{M}_1^2  (\mathcal{M}_1^2+2 )-3 )}{5 \mathcal{M}_1   (\mathcal{M}_1^2-1)^2},
\end{eqnarray}
where $\mathcal{M}_1$ is the upstream Mach number\footnote{Here we set $v_{\mathrm{thi},1} \sim c_{s1}$ in order to write $U_1/v_{\mathrm{thi},1} \sim \mathcal{M}_1$, where $c_{s1}$ is the upstream sound speed. An exact calculation only changes the end result by a factor of order unity. Moreover, the same factor also modifies the collisionless shock width (\ref{eq:widtlCollLess}). Therefore,  the critical plasma parameter $\Lambda_c$ defined by Eq. (\ref{eq:Lambda_c}) for the collisional/collisionless transition, remains unchanged when considering $v_{\mathrm{thi},1} \sim c_{s1}$.}. We eventually obtain the following limits for the shock width $\ell$,
\begin{equation}\label{eq:widthBGK_OK}
\ell = \lambda_{\mathrm{mfp},1}  \times \left\{\begin{array}{r}
                                                                  \frac{12}{5}(\mathcal{M}_1-1)^{-2} ~~\mathrm{for}~~\mathcal{M}_1 \sim 1, \\
                                                                   \mathcal{M}_1~~\mathrm{for}~~\mathcal{M}_1 \rightarrow \infty.
                                                                           \end{array} \right.
\end{equation}

\begin{figure}
\begin{center}
 \includegraphics[width=0.6\textwidth]{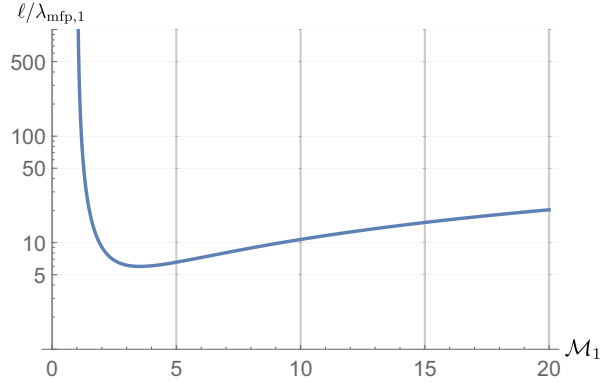}
\end{center}
\caption{Shock front thickness in units of the mean-free-path, $\ell/\lambda_{\mathrm{mfp},1}$, in terms of the upstream Mach number $\mathcal{M}_1$.}\label{widthcoll}
\end{figure}

The function $\ell/\lambda_{\mathrm{mfp},1}$ is plotted in Figure \ref{widthcoll} in terms of the Mach number. It reaches a minimum for $\mathcal{M}_1=3.53$ with $\ell/\lambda_{\mathrm{mfp},1}=6$. Such a ``U'' shape has also been found in \cite{Tidman1958}. We shall further comment on \cite{Tidman1958} in Section \ref{sec:tid58}.

\section{Collisionless regime}\label{sec:coll-less}
Our expression (\ref{eq:widthBGK_OK}) of the shock width cannot be used to bridge all the way to collisionless shocks since it has been derived from the kinetic equation (\ref{eq:kujl}) without the $(\partial F/\partial t)_w$ collision term. Yet,  collisionless shocks are sustained by the mechanism described by this very term.

\cite{Tidman1967} treated the problem of a collisionless shock by setting $(\partial F/\partial t)_c=0$ in Eq. (\ref{eq:kujl}) and considering the quasi-linear operator for $(\partial F/\partial t)_w$. The Mott-Smith \emph{ansatz} was also implemented in this study. \cite{Tidman1967} could not derive an equation of the form (\ref{eq:CollOK}) allowing to extract an analytical shock profile. Further analysis in \cite{Biskamp1969} and \cite{Tidman1969} concluded that the quasi-linear formalism is not non-linear enough to fully render a  shock.

Yet, \cite{Tidman1967} could derive the following estimate of the width of the front,
 \begin{eqnarray}\label{eq:A}
     \ell &=& A\frac{U_1}{\omega_{pi,1}} = A \frac{U_1}{v_{\mathrm{thi},1}}  \frac{v_{\mathrm{thi},1}}{\omega_{pi,1}} \nonumber \\
          &=& A \mathcal{M}_1  \lambda_{\mathrm{Di},1},
 \end{eqnarray}
 where $\lambda_{\mathrm{Di},1}$ is the upstream ionic Debye length and $A$ is a  parameter expected to be of order $\mathcal{O}(10)$. We can eventually cast this result under the form,
\begin{equation}\label{eq:widtlCollLess}
 \ell = A \mathcal{M}_1 \frac{\ln \Lambda}{\Lambda} \lambda_{\mathrm{mfp,1}},
\end{equation}
where $\Lambda$ is the plasma parameter already introduced in Eq. (\ref{GammaTid}) and we have used (\cite{fitzpatrick2014}, p. 10),
\begin{equation}\label{mfp_LD}
\lambda_{\mathrm{mfp,1}} = \frac{\Lambda}{\ln \Lambda}\lambda_{\mathrm{Di},1}.
\end{equation}

Notably,  \cite{Tidman1967} only addressed high Mach numbers turbulent shocks triggered by electrostatic instabilities. The forthcoming bridging between the 2 regimes is therefore only valid for such shocks. Weibel shocks sustained by electromagnetic instabilities are therefore excluded \citep{Stockem2014NatSR,Ruyer2017}.

\begin{figure}
\begin{center}
 \includegraphics[width=0.6\textwidth]{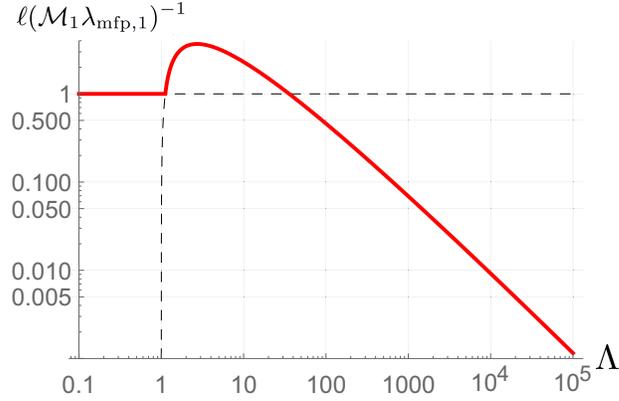}
\end{center}
\caption{Plot of $\ell (\mathcal{M}_1 \lambda_{\mathrm{mfp},1})^{-1}$ in terms of $\Lambda$. For a collisional plasma with small $\Lambda$, the front thickness $\ell$ is given by Eq. (\ref{eq:widthBGK_OK}). The collisionless thickness is given by Eq. (\ref{eq:widtlCollLess}). The width of the front for any plasma parameter $\Lambda$ is  given by the red curve. Only valid for strong shock (see end of Section \ref{sec:coll-less}).}\label{width}
\end{figure}

\begin{figure}
\begin{center}
 \includegraphics[width=0.6\textwidth]{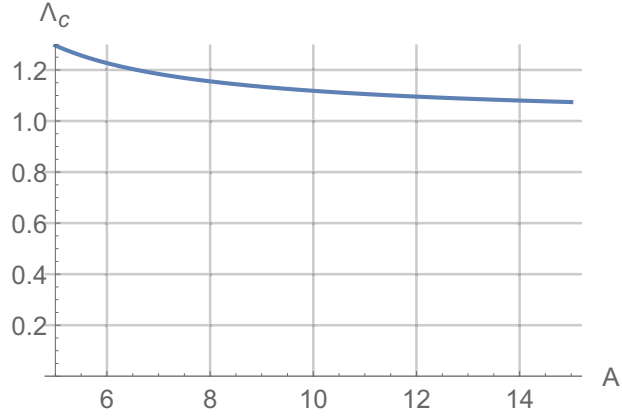}
\end{center}
\caption{Value of $\Lambda_c$ for which Eqs. (\ref{eq:widthBGK_OK} \& \ref{eq:widtlCollLess}) intersect, in terms of $A$ defined through Eq. (\ref{eq:A}).}\label{NcA}
\end{figure}

\section{Bridging  between the 2 regimes}\label{sec:bridge}
Figure \ref{width} shows the collisional and collisionless expressions of $\ell (\mathcal{M}_1 \lambda_{\mathrm{mfp},1})^{-1}$ from Eqs. (\ref{eq:widthBGK_OK}, \ref{eq:widtlCollLess}). For upstream Mach number $\mathcal{M}_1 >$ a few (4-5), these 2 expressions intersect for a critical plasma parameter $\Lambda_c$ defined by,
\begin{equation}\label{eq:Lambda_c}
\lambda_{\mathrm{mfp},1} \mathcal{M}_1 = A \mathcal{M}_1 \frac{\ln \Lambda_c}{\Lambda_c} \lambda_{\mathrm{mfp,1}} ~~ \Rightarrow ~~
  1 = A \frac{\ln \Lambda_c}{\Lambda_c},
\end{equation}
fulfilled for $\Lambda_c \sim 1.12$ and then for $\Lambda_c \sim 35$ (for $A=10$).

For $\Lambda < 1.12$, the upstream is strongly coupled, that is, collisional, and the width of the front will be given by the collisional result (\ref{eq:widthBGK_OK}). For $\Lambda > 1.12$, the upstream is weakly coupled, that is, collisionless, and the relevant front width is therefore the collisionless result (\ref{eq:widtlCollLess}). Hence, the larger value of $\Lambda_c \sim 35$ where the 2 expressions intersect again is not physically meaningful. For such values of $\Lambda$, the upstream is collisionless so that the collisionless result applies.

The transition between the 2 regimes occurs therefore for a critical plasma parameter $\Lambda_c = 1.12$, coinciding with the transition of the upstream from the strongly coupled/collisional regime, to the weakly coupled/collisionless regime. Although this value of $\Lambda_c$ has been computed for $A=10$, Figure \ref{NcA} shows it is poorly sensitive to $A$ as long as $A=\mathcal{O}(10)$.

Note that this value of $\Lambda_c = 1.12$ is only indicative. For example, \cite{LeeMore1984} developed an electron conductivity model for \emph{dense} plasmas requiring  $\ln \Lambda \geq 2$, i.e, $\Lambda \geq e^2=7.39$. Therefore, while $\Lambda =35$  probably pertains to weakly collisional plasmas, the value $\Lambda_c = 1.12$ only gives a general idea of where the transition occurs.

The width of the front for any plasma parameter $\Lambda$ is eventually given by the red curve in Figure \ref{width}. Simply put, the nature of the shock is the same as the nature of the upstream. Both are collisional or collisionless together.

The non-monotonic behavior in the collisionless regime is just the consequence of the non-monotonic variation of the mean-free-path in terms of the plasma parameter. The function $g(x)=A\ln x/x$ reaches a max for $x=e$ with $g(e)=3.67$, still for $A=10$.

\section{Comparison with   \cite{Tidman1958} }\label{sec:tid58}
A calculation parallel to the one performed in Section \ref{sec:coll} for the collisional regime was achieved in \cite{Tidman1958}. However, as we show here,  the bridging it provides to the collisionless regime is inadequate.

For the ion distribution function $F$, \cite{Tidman1958} used the Fokker-Planck operator for $(\partial F/\partial t)_c$ in Eq. (\ref{eq:kujl}), set $(\partial F/\partial t)_w=0$, and found for strong shocks\footnote{See Eq. (6.6) of \cite{Tidman1958} where $V$ is the sound speed and $K$ is the Mach number.},
\begin{equation}
\ell_T = \alpha \frac{c_s^4}{N_{1,0}\Gamma}\mathcal{M}_1^4,
\end{equation}
where $\alpha=29.1/512\pi$ and $\Gamma=\frac{4\pi e^4}{m_i^2}  \ln \Lambda $. We can recast this result under the form,
\begin{equation}\label{eq:ellTid}
\ell_T =   4\pi\alpha ~ \lambda_{\mathrm{mfp,1}} \mathcal{M}_1^4   \sim   0.23 ~ \lambda_{\mathrm{mfp,1}} \mathcal{M}_1^4,
\end{equation}
where we have used Eq. (\ref{mfp_LD}).

As a consequence, bridging the collisional result of \cite{Tidman1958} with the collisionless result of \cite{Tidman1967}, that is, bridging Eq. (\ref{eq:ellTid}) with Eq. (\ref{eq:widtlCollLess}), implicitly defines a critical plasma parameter $\Lambda_c$ through,
\begin{equation}\label{eq}
\frac{4\pi\alpha}{A }  \mathcal{M}_1^3  = \frac{\ln \Lambda_c}{\Lambda_c} ,
\end{equation}
 yielding  a Mach number-dependent value of $\Lambda_c$ and having no solution if the left-hand-side is larger than the maximum of the right hand-side, that is, for $\mathcal{M}_1 > 2.51$ (considering $A=10$).

As opposed to that, the $\propto \mathcal{M}_1$ scaling of the collisional $\ell$ given by BGK-derived Eq. (\ref{eq:widthBGK_OK}) is essential to give a value of $\Lambda_c $ independent of the upstream Mach number $\mathcal{M}_1$, with a switch from the collisional to the collisionless regime when the upstream becomes collisionless.

\bigskip

We therefore find that BGK provides a better bridging to the collisionless regime than Fokker-Planck. \cite{Hazeltine1998} already noted the capacity of the BGK operator to behave adequately in the collisionless limit. Computing the moments of the kinetic equation with the BGK  operator, he could derive a \emph{non-local} expression of the heat flux in the collisionless regime, as expected when the mean-free-path becomes large \citep{Hammett1990,Hazeltine1998}. Indeed, the BGK operator was specifically designed to provide an operator capable of giving an adequate description of low-density plasmas \citep{Krook1954}.

The physical reason for the better behavior of the BGK operator when the mean free path becomes large could be that regardless of the mean free path, BGK assumes the equilibrium distribution function is a Maxwellian, since the collision term (\ref{eq:KrookColl}) vanishes for $F=N\phi$, where $\phi$ is a Maxwellian (see Eq. \ref{krookPhi}).

In contrast, the Fokker-Planck operator does not assume any a priori form of the equilibrium distribution function. It can even be used to prove that such a function is a Maxwellian. Yet, the collision rate is implicitly assumed large compared to the dynamic terms $v/L$ in the Fokker-Planck equation (\cite{Kulsrud2005}, p. 213) since  the derivation of the Fokker-Planck operator involves a Taylor expansion in time, implicitly assuming collisions are frequent enough (\cite{Kulsrud2005}, Eq. 29-30, p. 204 or \cite{Chandrasekhar1943},  \S II.4).

Therefore, when collisions become scarce, the BGK formalism keeps forcing, by design, a Maxwellian equilibrium, while Fokker-Planck progressively loses validity.

\section{Conclusion}
We propose a bridging between collisional and collisionless shocks. The collisional ``leg'' is worked out using the Moot-Smith \emph{ansatz} \citep{MottSmith1951} with the ``full'' BGK collision term which behaves correctly in the large mean-free-path limit \citep{Krook1954,Krook1956,Hazeltine1998}. The collisionless part is from \cite{Tidman1967}, valid for strong turbulent electrostatic shocks.

The result makes perfect physical sense. As long as the upstream is strongly coupled, that is, collisional with $\Lambda \lesssim 1.12$, the strong shock is collisional with a front thickness $\sim \mathcal{M}_1 \lambda_{\mathrm{mfp},1}$ given by Eq. (\ref{eq:widthBGK_OK}). From $\Lambda \gtrsim 1.12$, the shock switches to the collisionless regime, with a front thickness $\ell \sim \mathcal{M}_1 \lambda_{\mathrm{mfp},1}\ln\Lambda /\Lambda$, given by Eq. (\ref{eq:widtlCollLess}).

We show that the BGK treatment of the collisional regime provides a better bridge to the collisionless regime than the Fokker-Planck model. Nevertheless, a confusing feature remains: in the collisional limit, one would expect the BGK and the Fokker-Planck treatments to merge. Yet, they don't, as evidenced by their different $\mathcal{M}_1$ scaling for the strong shock width ($\propto\mathcal{M}_1$ for BGK vs. $\propto \mathcal{M}_1^4$ for Fokker-Planck). The reason for this could be that the collision frequency used in BGK (Eq. \ref{eq:nuBGK}) does not depend on the particle velocity. However, this is still unclear to us.

A smoother transition between the 2 regimes could be assessed from Eq. (\ref{eq:kujl}) considering both $(\partial F/\partial t)_c$ and $(\partial F/\partial t)_w$ at once, whereas we here switched them on and off according to the regime considered. The Mott-Smith \emph{ansatz} could still be applied, while using BGK  for $(\partial F/\partial t)_c$ and the operator proposed by \cite{Dupree1966} (as suggested in \cite{Tidman1967})  or  \cite{BaalrudPoP2008}, for $(\partial F/\partial t)_w$.

Although the present theory is formally restricted to high Mach number, un-magnetized, electrostatic shocks, it may help understand how the value of $\Lambda \sim 10^{10}$ observed in the solar wind (see for example \cite{fitzpatrick2014}, p. 8) yields an earth bow shock thickness orders of magnitude shorter than the mean-free-path.

\section{Acknowledgments}
A.B. acknowledges support by grants  ENE2016-75703-R from the Spanish Ministerio de Econom\'{\i}a y Competitividad and SBPLY/17/180501/000264 from the Junta de Comunidades de Castilla-La Mancha.

A. P. acknowledges support from the European Research Council via ERC consolidating grant \#773062 (acronym O.M.J.).

Thanks are due to Anatoly Spitkovsky, Bill Dorland, Ian Hutchinson, Ellen Zweibel and Richard Halzeltine for valuable inputs.

\appendix

\section{Proof of Eq. (\ref{eq:CollOK})}\label{ap:proofBGK}
Equation (\ref{eq:CollOK}) is the $v_y^2$ moment of Eq. (\ref{eq:Krook3}). The left-hand-side is calculated in \cite{Tidman1958}. Note that the term proportional to $E_x$ vanishes in this moment. We only detail here the calculation proper to the present work, that is, that of the right-hand-side. For this we need $\Phi$, hence $\mathbf{q}$ and $T$ defined by Eqs. (\ref{krookN}-\ref{krookT}).

According to Eq. (\ref{krookq}), $\mathbf{q}$  is given by,
\begin{equation}
  \mathbf{q}  = \frac{1}{N}\int \mathbf{v} F d^3v = \frac{1}{n_1 + n_2}\int \mathbf{v} F d^3v .
\end{equation}
Since $F$ is the sum of 2 drifting Maxwellians given by Eq. (\ref{eq:amsatz}), we find for $\mathbf{q} $,
\begin{equation}
  \mathbf{q}  =  \frac{n_1 \mathbf{U}_1 + n_2 \mathbf{U}_2}{n_1 + n_2} \equiv q ~ \mathbf{e}_x,
\end{equation}
where $\mathbf{e}_x$ is the unit vector of the $x$ axis. For $T$ we then get from (\ref{krookT})\footnote{The factor 2 in the second term of Eq. (\ref{fact2}) comes from $\int v_z^2F = \int v_y^2F $.},
\begin{eqnarray}
\frac{3  k_BT }{m_i}  &=&  \frac{1}{n_1 + n_2}\int (\mathbf{v}-\mathbf{q} )^2 F d^3v \nonumber \\
 &=&  \frac{1}{n_1 + n_2}\int [(v_x-q )^2 + v_y^2 + v_z^2] F d^3v \nonumber \\
 &=&  \frac{1}{n_1 + n_2}\int (v_x-q )^2  F d^3v + \frac{2}{n_1 + n_2}\int v_y^2 F d^3v \label{fact2} \\
  &=&  \frac{1}{n_1 + n_2}\int (v_x-q )^2  F d^3v +  \frac{2  (k_BT_1 n_1+k_BT_2 n_2 )}{m_i  ( n_1+n_2 )} \nonumber \\
 \Rightarrow  k_BT    &=&  \frac{n_1 k_BT_1 +n_2 k_BT_2 }{n_1+n_2}
                                           +\frac{n_1 n_2}{3 (n_1+n_2 )^2}m_i  (U_1-U_2 )^2.
\end{eqnarray}
Let us now write explicitly the $v_y^2$ moment of the right-hand-side ($rhs$) of Eq. (\ref{eq:Krook3}),
\begin{equation}\label{eq:rhs}
rhs = \frac{1}{ \sigma }\int v_y^2 (-N F + N^2 \Phi )d^3v
 = \underbrace{\frac{ N^2}{ \sigma }\int v_y^2 \Phi d^3v}_{\mathbf{1}} - \underbrace{\frac{ N}{ \sigma }\int v_y^2 F d^3v}_{\mathbf{2}}.
\end{equation}
From (\ref{krookN}) we see $N$ does not depend on $\mathbf{v}$. It can therefore be taken out of the integrals.

Computing \textbf{2} we find,
\begin{eqnarray}
\frac{ N}{ \sigma }\int v_y^2 F d^3v
        &=& \frac{ n_1 + n_2}{ \sigma } \left(  n_1   \frac{  k_BT_1}{m_i}  +  n_2   \frac{k_BT_2}{m_i} \right), \nonumber \\
        &=& \frac{  k_B}{m_i\sigma  }( n_1 + n_2   ) ( n_1 T_1 + n_2 T_2   ).
\end{eqnarray}
Then we compute \textbf{1}.
\begin{eqnarray}
 \frac{ N^2}{ \sigma }\int v_y^2 \Phi d^3v
  &=& \frac{ (n_1 + n_2)^2}{ \sigma }\int v_y^2 \left( \frac{m_i}{2\pi k_BT } \right)^{3/2}\exp\left( -\frac{m_i}{2  k_BT }(\mathbf{v}-\mathbf{q} )^2 \right)d^3v, \nonumber \\
   &=& \frac{ (n_1 + n_2)^2}{ \sigma }\left( \frac{m_i}{2\pi k_BT } \right)^{3/2}\underbrace{\int v_y^2\exp\left( -\frac{m_i}{2  k_BT }(\mathbf{v}-\mathbf{q} )^2 \right)d^3v}_{\mathbf{3}}. \nonumber
\end{eqnarray}
For \textbf{3} we get,
\begin{equation}
\mathbf{3} =   (2\pi)^{3/2} \left(\frac{k_BT }{m_i}\right)^{5/2},
\end{equation}
so that \textbf{1} gives,
\begin{eqnarray}
  \mathbf{1} &=& \frac{ (n_1 + n_2)^2}{ \sigma }\left( \frac{m_i}{2\pi k_BT } \right)^{3/2}(2\pi)^{3/2} \left(\frac{k_BT }{m_i}\right)^{5/2} \nonumber\\
  &=& \frac{ (n_1 + n_2)^2}{ \sigma } \frac{k_BT }{m_i}  .
\end{eqnarray}
Finally, Eq. (\ref{eq:rhs}) simplifies nicely and reads,
\begin{eqnarray}
rhs &=& \frac{ (n_1 + n_2)^2}{\sigma } \frac{k_BT }{m_i} - \frac{k_B}{m_i\sigma }( n_1 + n_2) ( n_1 T_1 + n_2 T_2) , \nonumber\\
  &=& \frac{1}{3\sigma } n_1 n_2  (U_1-U_2 )^2,
\end{eqnarray}
in agreement with the right-hand-side of Eq. (\ref{eq:CollOK}).

\section{Derivation of the density profiles (\ref{eq:profile}) from Eq. (\ref{eq:CollOK})}\label{ap:profile}
Let us define $\alpha,\beta,\gamma$ from Eq. (\ref{eq:CollOK}) by,
\begin{equation}\label{eq:alphabeta}
\underbrace{U_1\frac{k_BT_1}{m_i}}_{\alpha}\frac{\partial n_1}{\partial x} + \underbrace{U_2\frac{k_BT_2}{m_i}}_{\beta}\frac{\partial n_2}{\partial x}
                 = \underbrace{\frac{1}{3\sigma } (U_1-U_2 )^2}_\gamma n_1 n_2.
\end{equation}
Consider now the matter conservation equation obtained equating the $v_x$ moments of (\ref{eq:amsatz}) between any $x$ and $x=+\infty$,
\begin{equation}\label{eq:consermatt}
n_1(x)U_1 +  n_2(x)U_2 = N_{1,0}U_1 .
\end{equation}
Differentiate with respect to $x$ gives,
\begin{equation}
\frac{\partial n_1(x)}{\partial x}U_1 +  \frac{\partial n_2(x)}{\partial x}U_2 = 0 ~~ \Rightarrow ~~
                                          \frac{\partial n_2(x)}{\partial x} = -\frac{\partial n_1(x)}{\partial x}\frac{U_1}{U_2},
\end{equation}
and use the result to eliminate $\partial n_2/\partial x$ in  (\ref{eq:alphabeta}),
\begin{equation}
\frac{\partial n_1}{\partial x}\left( \alpha - \beta \frac{U_1}{U_2}\right) = \gamma n_1 n_2 ~~ \Rightarrow ~~
\frac{\partial n_1}{\partial x}\frac{1}{n_1 n_2} = \frac{\gamma}{\alpha - \beta \frac{U_1}{U_2}}.
\end{equation}
Making now use again of the conservation equation (\ref{eq:consermatt}) to write,
\begin{equation}
 n_2 = (N_{1,0} - n_1)\frac{U_1}{U_2},
\end{equation}
one gets,
\begin{equation}
\frac{U_2}{U_1}\frac{\partial n_1}{\partial x}\frac{1}{n_1 (N_{1,0} - n_1)}
=\frac{U_2}{U_1}\frac{\partial n_1}{\partial x}\frac{1}{N_{1,0}} \left(  \frac{1}{n_1} +  \frac{1}{N_{1,0} - n_1 }  \right)
 = \frac{\gamma}{\alpha - \beta \frac{U_1}{U_2}}.
\end{equation}
We eventually obtain,
\begin{equation}\label{eq:pattern}
\frac{\partial n_1}{\partial x} \left(  \frac{1}{n_1} +  \frac{1}{N_{1,0} - n_1 }  \right)
 = N_{1,0}\frac{\gamma}{\alpha - \beta \frac{U_1}{U_2}} \frac{U_1}{U_2}\equiv -\ell^{-1},
\end{equation}
where $\ell$ is the shock thickness since the solution accounting for the boundary conditions (\ref{eq:boudary}) is,
\begin{equation}\label{eq:n1pattern}
n_1(x) = N_{1,0} \frac{1}{1+e^{-x/\ell}}.
 \end{equation}
From (\ref{eq:consermatt}) one then obtains for $n_2(x)$,
\begin{equation}\label{eq:n2pattern}
n_2(x) = N_{2,0} \frac{e^{-x/\ell}}{1+e^{-x/\ell}}.
 \end{equation}

\section{Derivation of Eq. (\ref{eq:stage}) from Eq. (\ref{eq:ell_coll})}\label{ap:RH}
We first cast Eq. (\ref{eq:ell_coll}) under the form,
\begin{equation}
\ell=3\sigma \frac{  k_BT_1(1-T_2/T_1) (U_2/U_1)U_1  }{ N_{1,0}  m_i U_1^2(1-U_2/U_1)^2}.
\end{equation}
We then use the RH jump conditions (see for example \cite{fitzpatrick2014} p. 216, or \cite{thorne2017modern} p. 905),
\begin{equation}
  \left(\frac{U_2}{U_1}\right)^{-1} =  \frac{N_{2,0}}{N_{1,0}} = \frac{\gamma + 1}{\gamma - 1 + 2 \mathcal{M}_1^{-2}},
\end{equation}
and,
\begin{equation}
\frac{T_2}{T_1} = \frac{P_2}{P_1}\frac{N_{1,0}}{N_{2,0}}
\end{equation}
with,
\begin{equation}
  \frac{P_2}{P_1} = \frac{2\gamma\mathcal{M}_1^2 - \gamma + 1 }{\gamma + 1 }.
\end{equation}
Substituting these ratios and setting,
\begin{equation}
\mathcal{M}_1^2 =\frac{U_1^2}{\gamma P_1/N_{1,0}}
\end{equation}
we get to Eq. (\ref{eq:stage}) with $\gamma=5/3$.


\end{document}